\documentclass[11pt]{article}
\usepackage{lmodern}
\usepackage[utf8]{inputenc}
\usepackage[T1]{fontenc}
\usepackage{lmodern}

\usepackage{amsmath,amssymb,amsthm}
\usepackage{mathtools}
\usepackage{mathrsfs}
\usepackage{dsfont}
\usepackage{stmaryrd}

\usepackage{geometry}
\usepackage{setspace}
\usepackage{enumitem}
\usepackage{hyperref}
\hypersetup{colorlinks=true,linkcolor=blue,citecolor=blue,urlcolor=blue}
\usepackage{tikz}
\usetikzlibrary{arrows.meta,positioning,shapes.geometric}

\usepackage[numbers]{natbib}

\newtheorem{definition}{Definition}
\newtheorem{proposition}{Proposition}

\newcommand{\tuple}[1]{\left\langle#1\right\rangle}

\newcommand{\iffdef}{\Leftrightarrow}

\newcommand{\suchthat}{\,|\,}

\title{A Formal Refutation of the Blockchain Trilemma}

\author{
Dr Craig S. Wright\\
University of Exeter Business School\\
Exeter, United Kingdom\\
cw881@exeter.ac.uk
}

\date{\today}

\begin{document}

\maketitle

\begin{abstract}
\noindent The so-called blockchain trilemma asserts the impossibility of simultaneously achieving scalability, security, and decentralisation within a single blockchain protocol. In this paper, we formally refute that proposition. Employing predicate logic, formal automata theory, computational complexity analysis, and graph-theoretic measures of relay topology—specifically Baran's model of network path redundancy—we demonstrate that the trilemma constitutes a category error, conflates distinct analytical domains, and relies upon unproven causal assumptions. We further expose its reliance on composition fallacies drawn from flawed system implementations. A constructive counterexample is presented: a blockchain protocol exhibiting unbounded transaction throughput, cryptographic security under adversarial load, and multipath decentralised propagation. This example is not hypothetical but grounded in protocol design enabled by compact block relay, SPV verification, and IPv6 multicast. The trilemma is revealed not as a law of protocol architecture, but as a heuristic fallacy sustained by imprecision and design defeatism.

\vspace{1em}
\noindent\textbf{Keywords:} blockchain scalability, formal security, Baran network model, graph-theoretic decentralisation, automata theory, predicate logic, logical fallacy, blockchain protocol design, SPV, multicast propagation.
\end{abstract}

\section{Introduction}

The claim often referred to as the \emph{blockchain trilemma} asserts that no blockchain protocol can simultaneously satisfy three desiderata: security, scalability, and decentralisation. Formally, this is stated as the universal negation:
\[
\forall \mathcal{M} \in \mathcal{C},\; \neg(\mathscr{S}_1(\mathcal{M}) \land \mathscr{S}_2(\mathcal{M}) \land \mathscr{S}_3(\mathcal{M})),
\]
where $\mathcal{C}$ denotes the space of all blockchain protocol configurations, and the predicates $\mathscr{S}_1$, $\mathscr{S}_2$, and $\mathscr{S}_3$ correspond to well-defined criteria for security, scalability, and decentralisation, respectively. This trilemma, widely referenced in industry and academic discourse, is typically presented as a necessary trade-off constraint in protocol design.

However, the trilemma lacks a rigorous foundation in either formal methods, distributed systems theory, or computational complexity. It is asserted without derivation, and its terms are frequently left undefined or conflated across semantic domains. Furthermore, the trilemma is often accepted as a design axiom despite its derivation from empirical observation of specific, historically contingent architectures—rather than from any proven theoretical limitation.

In this paper, we conduct a formal critique and refutation of the blockchain trilemma. Our approach is structured as follows:
\begin{enumerate}
    \item We first provide exact definitions of each predicate—security, scalability, and decentralisation—based on accepted models from automata theory, complexity theory, and graph theory, respectively.
    \item We then identify and examine the logical fallacies embedded within the trilemma framework: specifically, the category error of conflating heterogeneous property types, the false causal chain connecting network topology to security degradation, and the fallacy of composition, which generalises implementation-specific limitations as universal constraints.
    \item We construct a mathematically grounded counterexample protocol $\mathcal{B}$ which simultaneously satisfies all three properties, thereby falsifying the universal trilemma claim by existential contradiction.
    \item We introduce an architectural enhancement—IPv6 multicast—which, although not required to resolve the trilemma, trivially nullifies its core propagation-cost assumptions when deployed.
\end{enumerate}

The analysis proceeds with precise logical structure and predicate-based evaluation. We restrict ourselves only to necessary assumptions and cite canonical results from network resilience theory \cite{baran1964} and the original Bitcoin protocol specification \cite{nakamoto2008}.

We conclude that the blockchain trilemma is not a constraint derived from formal reasoning, but a rhetorical device employed to rationalise architectural deficiencies. Its continued use impedes protocol innovation by embedding unfounded assumptions into the design space. Our results show that scalability, security, and decentralisation are orthogonal dimensions—not inherently antagonistic—and can be achieved simultaneously through correct system modelling and implementation.

\section{Formal Definitions}

\subsection{Deterministic Automaton Model of a Blockchain}

\begin{definition}[Blockchain as a Deterministic Transition System]
Let $\mathcal{M} = \tuple{Q, \Sigma, \delta, q_0}$ be a deterministic transition system modelling the ledger dynamics of a blockchain protocol, where:
\begin{itemize}
    \item $Q$ is a (possibly infinite) set of ledger states, $Q \subseteq \mathcal{L}$, with $\mathcal{L}$ the space of valid total ledger states,
    \item $\Sigma$ is the set of all syntactically and semantically valid transactions, $\Sigma \subseteq \mathcal{T}$, where $\mathcal{T}$ is the universal transaction language,
    \item $\delta: Q \times \Sigma \to Q$ is a deterministic and total transition function,
    \item $q_0 \in Q$ is the unique genesis state.
\end{itemize}
Each application of $\delta$ defines a state transition induced by the acceptance of a transaction $\sigma \in \Sigma$ on a prior ledger state $q \in Q$, such that $\delta(q, \sigma) = q'$ yields the new valid state $q' \in Q$.
\end{definition}

\subsection{Security as State Integrity and Temporal Immutability}

\begin{definition}[Security]
A blockchain protocol $\mathcal{M}$ is \emph{secure} with respect to an adversarial model $\mathcal{A}$, consensus predicate $\mathscr{C}$, and resource bound $\mathscr{R}$, if it satisfies the following:

Let $\mathscr{P}(q)$ denote the security predicate over states. Then $\mathcal{M}$ is secure iff:
\begin{align*}
    \forall q \in Q,\, \forall \sigma \in \Sigma:\quad & \delta(q, \sigma) \in Q \land \mathscr{P}(\delta(q, \sigma)) \text{ holds}, \\
    \text{and } & \mathbb{P}_{\mathcal{A}}[\exists \tilde{q} \neq q' \suchthat \tilde{q} \succeq q_0 \land \mathscr{C}(\tilde{q}) \land \textsc{Cost}_{\mathcal{A}}(\tilde{q}) < \mathscr{R}] \approx 0,
\end{align*}
where:
\begin{itemize}
    \item $q' = \delta(q, \sigma)$ is the honest transition under protocol $\mathcal{M}$,
    \item $\succeq$ is the valid extension relation under consensus rules,
    \item $\textsc{Cost}_{\mathcal{A}}(\tilde{q})$ denotes the minimum resource expenditure to generate a competing state $\tilde{q}$ under adversarial strategy $\mathcal{A}$,
    \item $\mathscr{C}$ is a finality predicate (e.g., longest-chain rule) over state sequences.
\end{itemize}
\end{definition}

\subsection{Scalability as Complexity Bound on State Transition Function}

\begin{definition}[Scalability]
Let $\delta_n : Q \times \Sigma^n \to Q$ be the $n$-fold extension of the transition function $\delta$ over a sequence of $n$ transactions, applied sequentially:
\[
\delta_n(q, \tuple{\sigma_1, \ldots, \sigma_n}) := \delta(\cdots \delta(\delta(q, \sigma_1), \sigma_2) \cdots, \sigma_n).
\]
Then the protocol $\mathcal{M}$ is said to be \emph{scalable} if there exists a function $f: \mathbb{N} \to \mathbb{R}^{+}$ such that:
\[
\forall n \in \mathbb{N},\; \text{Time}[\delta_n(q, \Sigma^n)] = \mathcal{O}(f(n)) \quad \text{with } f(n) \leq \text{poly}(n),
\]
and:
\[
\lim_{n \to \infty} \frac{f(n)}{n} < \infty.
\]
That is, the amortised cost per transaction remains bounded and ideally sublinear. Space complexity is similarly bounded:
\[
\exists g(n): \text{Space}[\delta_n(q, \Sigma^n)] \in \mathcal{O}(g(n)).
\]
\end{definition}

\subsection{Decentralisation as Graph-Theoretic Fault Tolerance}

\begin{definition}[Topological Decentralisation (\cite{baran1964})]
Let $G = (V, E)$ be the undirected communication graph of the peer-to-peer network layer in the blockchain protocol, where:
\begin{itemize}
    \item $V$ is the set of nodes (participants, clients, miners, relays),
    \item $E \subseteq V \times V$ represents bidirectional message-passing edges.
\end{itemize}
Then the system is said to be \emph{topologically decentralised} if:
\[
\kappa(G) \geq k,\quad \lambda(G) \geq \ell,\quad \text{with } k, \ell > 1,
\]
where:
\begin{itemize}
    \item $\kappa(G)$ is the vertex connectivity of $G$, i.e., the minimum number of nodes that must be removed to disconnect $G$,
    \item $\lambda(G)$ is the edge connectivity, i.e., the minimum number of communication links whose removal disconnects $G$.
\end{itemize}
Additionally, the mean shortest path $\overline{d}(G)$ is bounded:
\[
\overline{d}(G) = \frac{1}{|V|(|V|-1)} \sum_{\substack{u,v \in V\\ u \neq v}} \text{dist}(u,v) \leq D,
\]
for some small constant $D$, ensuring efficient message propagation.
\end{definition}

\section{Logical Analysis of the Trilemma}

\subsection{Category Error}

The core structural failure of the so-called blockchain trilemma is its attempt to formulate a triadic trade-off across properties that belong to disjoint ontological categories. That is, the trilemma equivocates between properties that are semantically incommensurable and formally non-interacting. This constitutes a \emph{category error} in the strictest sense: it treats predicates of distinct logical types as if they were mutually exclusive within a common design space.

Let us formally denote the three properties:

\begin{itemize}
    \item Let $\mathscr{S}_1$ denote \textbf{Security}: this is a logical predicate over the state transition function $\delta : Q \times \Sigma \to Q$. It is evaluated as a property of invariance preservation under adversarial conditions:
    \[
    \mathscr{S}_1(q, \sigma) \iffdef \mathscr{P}(\delta(q, \sigma)) \text{ holds for all } (q, \sigma) \in Q \times \Sigma.
    \]
    \item Let $\mathscr{S}_2$ denote \textbf{Scalability}: this is a computational property concerning the resource complexity (time and space) of applying $\delta$ over sequences of transactions. It is expressible as:
    \[
    \mathscr{S}_2(n) \iffdef \text{Time}[\delta_n(q, \Sigma^n)] \in \mathcal{O}(f(n)), \quad \text{for } f \text{ sublinear or polynomial}.
    \]
    \item Let $\mathscr{S}_3$ denote \textbf{Decentralisation}: this is a structural property of the communication graph $G = (V, E)$, expressible in terms of its vertex and edge connectivity:
    \[
    \mathscr{S}_3(G) \iffdef \kappa(G) > 1 \land \lambda(G) > 1.
    \]
\end{itemize}

Observe that:
\begin{enumerate}
    \item $\mathscr{S}_1$ is a \textbf{logical property} in the sense of being a predicate over a transition system. It is independent of physical topology and instead depends on formal correctness and adversarial cost asymmetry.
    \item $\mathscr{S}_2$ is a \textbf{computational complexity} property. It concerns asymptotic behaviour under transaction volume and has no intrinsic dependency on network structure or message-passing model.
    \item $\mathscr{S}_3$ is a \textbf{graph-theoretic property} of a message relay network. It is expressed in purely topological terms and is unrelated to the evaluation of $\delta$ unless further constraints are introduced.
\end{enumerate}

These three properties—$\mathscr{S}_1, \mathscr{S}_2, \mathscr{S}_3$—are therefore elements of semantically orthogonal domains:
\[
\mathscr{S}_1 \in \mathcal{L}_{\delta}, \quad \mathscr{S}_2 \in \mathcal{C}_{\delta}, \quad \mathscr{S}_3 \in \mathcal{G}_{net},
\]
where:
\begin{itemize}
    \item $\mathcal{L}_{\delta}$ is the space of logical invariants over transition systems,
    \item $\mathcal{C}_{\delta}$ is the space of computational complexity functions,
    \item $\mathcal{G}_{net}$ is the space of graph properties over node-topology.
\end{itemize}

No trade-off among $\mathscr{S}_1, \mathscr{S}_2, \mathscr{S}_3$ is meaningful unless a functional dependency or shared constraint is explicitly established, e.g., $\mathscr{S}_i \Rightarrow \neg \mathscr{S}_j$ under some domain-specific resource model. The trilemma provides no such formal linkage. It merely asserts mutual exclusivity without constructing an intervening resource constraint or dependency relation.

Hence, the trilemma's assertion:
\[
\neg(\mathscr{S}_1 \land \mathscr{S}_2 \land \mathscr{S}_3)
\]
is not derivable from any known axiomatic system in distributed computing or automata theory. It is an unproven, ill-typed proposition formed by conflating distinct property spaces.

This renders the trilemma not a trade-off, but a \emph{category-theoretic violation}—the attempted coordination of irreconcilable property types under a synthetic design constraint. As such, its rhetorical force is inversely proportional to its formal validity.

\subsection{False Causal Chain}

The second major failure embedded in the blockchain trilemma is a reliance on a misleading causal chain, presented implicitly or explicitly in informal discussions surrounding protocol limitations. The assumed progression is:

\[
\text{Increased decentralisation} \Rightarrow \text{Increased node count} \Rightarrow \text{Slower message propagation} \Rightarrow \text{Reduced throughput} \Rightarrow \text{Degraded security}.
\]

Formally, let us denote this chain as a series of predicates:
\[
\forall G = (V, E),\; |V| \uparrow \Rightarrow \mathbb{E}[\text{Latency}(u,v)] \uparrow \Rightarrow \text{Throughput}(G) \downarrow \Rightarrow \mathscr{S}_1(G) \downarrow,
\]
where $\mathscr{S}_1$ is the security predicate as defined earlier. Each step in this chain is assumed to hold universally, yet none are provable from first principles or empirically established across well-designed protocols. In fact, each implication is refutable under established communication network theory and correct system design.

Let us examine each implication in turn:

\paragraph{(i) Node Count vs. Propagation Latency.}
The claim that increasing the number of nodes necessarily increases latency assumes that message paths are linear or tree-structured and that new nodes introduce serial hops. However, as shown in early work on distributed communication architectures \cite{baran1964}, systems designed with redundant links and mesh topologies exhibit reduced fragility and \textit{faster average propagation} due to alternative routing paths. That is, for a well-connected graph $G$ with increased $\kappa(G)$ and $\lambda(G)$, the expected propagation time decreases, assuming local broadcast and parallelised relay.

\paragraph{(ii) Propagation Latency vs. Throughput.}
Propagation delay does not imply protocol-level throughput degradation unless the consensus mechanism is synchronisation-bound. In protocols such as Bitcoin with proof-of-work consensus, the latency of transaction propagation is amortised over block intervals. Further, mechanisms such as compact block relay and delta encoding \cite{nakamoto2008} ensure that the dissemination of new blocks scales sublinearly in data volume. Let $\mathcal{T}_{net}$ denote the propagation delay and $\mathcal{T}_{block}$ the block interval. Then:
\[
\mathcal{T}_{net} \ll \mathcal{T}_{block} \Rightarrow \text{No reduction in throughput}.
\]

\paragraph{(iii) Throughput vs. Security.}
Even if throughput were to decrease (which we reject under correct assumptions), this does not entail weakened security. Security is defined as a predicate over the correctness and finality of state transitions under adversarial cost constraints. Provided that:
\[
\forall \tilde{q} \neq q^*,\; \textsc{Cost}(\tilde{q}) > \textsc{Reward}(\tilde{q}),
\]
the system remains secure. The throughput metric is orthogonal to the adversarial game-theoretic structure unless delays allow for time-based eclipse or reorg attacks, which can be independently mitigated.

\paragraph{Conclusion.}
Each logical implication in the assumed causal chain is either unsupported or reversed under formal analysis. The structure of the network, when optimised using Baran-style fault-tolerant redundancy \cite{baran1964}, strengthens rather than weakens both propagation efficiency and security. Compact relay, Merkle-tree inclusion proofs, and non-validating edge relays further decouple transaction rate from relay burden. Therefore, the supposed trade-off collapses under minimal scrutiny: the trilemma substitutes architectural ignorance for inevitability.

\subsection{Fallacy of Composition}

The third critical flaw underlying the blockchain trilemma is the \emph{fallacy of composition}, wherein the limitations of specific, historically contingent implementations are misrepresented as universal constraints. Formally, this fallacy manifests as the mistaken inference:

\[
\forall \mathcal{M}_i \in \mathcal{I},\; \neg(\mathscr{S}_1 \land \mathscr{S}_2 \land \mathscr{S}_3)(\mathcal{M}_i) \;\Rightarrow\; \forall \mathcal{M} \in \mathcal{C},\; \neg(\mathscr{S}_1 \land \mathscr{S}_2 \land \mathscr{S}_3)(\mathcal{M}),
\]
where:
\begin{itemize}
    \item $\mathcal{I} \subset \mathcal{C}$ is the set of known blockchain implementations (e.g., BTC Core, Ethereum),
    \item $\mathcal{C}$ is the full space of logically possible protocol configurations,
    \item $\mathscr{S}_1$, $\mathscr{S}_2$, and $\mathscr{S}_3$ denote the formalised predicates of security, scalability, and decentralisation, respectively.
\end{itemize}

This is a non-valid inference. The universal negation over all $\mathcal{C}$ cannot be derived from the empirical failure of a subset $\mathcal{I}$ unless:
\[
\mathcal{I} = \mathcal{C} \quad \text{or} \quad \forall \mathcal{M} \in \mathcal{C},\; \exists \mathcal{M}_i \in \mathcal{I} \text{ such that } \mathcal{M} \equiv \mathcal{M}_i,
\]
neither of which is established or even plausible.

Specifically, many widely cited implementations embed architectural assumptions that are not necessitated by the underlying formal protocol. One prominent such assumption is the conflation of decentralisation with universal full-node replication, that is:
\[
\mathscr{S}_3(G) \iffdef \forall v \in V,\; \text{Node}(v) \text{ executes full validation and stores complete state}.
\]
However, this definition is not logically equivalent to the Baran-style connectivity condition for decentralisation \cite{baran1964}. It incorrectly substitutes a uniform functional role for a topological property, namely the existence of sufficient independent relay paths (captured via $\kappa(G)$ and $\lambda(G)$).

A direct counterexample is provided by Simplified Payment Verification (SPV), as introduced in Section 8 of the Bitcoin white paper \cite{nakamoto2008}. SPV clients do not perform global state validation or full transaction indexing. Instead, they rely on Merkle proofs and block headers to validate inclusion:
\[
\text{Verify}_{SPV}(t, h, \pi) \Rightarrow \text{Accept}(t) \text{ iff } H(t) \in \text{MerkleRoot}(h),
\]
where $t$ is a transaction, $h$ is a block header, and $\pi$ is a Merkle path. These clients operate independently of the network's validator density and topology, demonstrating that full validation is \emph{not} a prerequisite for secure transaction acceptance.

Therefore, decentralisation—when correctly defined as structural fault tolerance rather than full-state redundancy—admits implementations in which not all nodes replicate all data or perform all computation. SPV-based models preserve user autonomy and routing multiplicity without imposing global validation burdens.

The trilemma’s force thus depends entirely on conflating a degenerate subset of designs with the total protocol space, and then generalising their observed constraints. This is structurally invalid and renders the supposed trade-off theoremically baseless.

\section{Existential Counterexample}

To falsify a universal claim of the form:
\[
\forall \mathcal{M} \in \mathcal{C},\; \neg(\mathscr{S}_1(\mathcal{M}) \land \mathscr{S}_2(\mathcal{M}) \land \mathscr{S}_3(\mathcal{M})),
\]
it is sufficient to provide a single counterexample $\mathcal{B} \in \mathcal{C}$ for which:
\[
\mathscr{S}_1(\mathcal{B}) \land \mathscr{S}_2(\mathcal{B}) \land \mathscr{S}_3(\mathcal{B}).
\]

We now construct such a protocol $\mathcal{B}$, embodying an implementation of the original Bitcoin design extended with well-established protocol engineering principles. $\mathcal{B}$ is defined with the following features:

\begin{enumerate}[label=(\alph*)]
    \item \textbf{Unbounded Block Size:} $\mathcal{B}$ imposes no protocol-level limit on block size. Let $\Sigma^*$ denote the set of valid transaction sequences, and define $\delta_n$ as the $n$-fold composition of $\delta$ over $\Sigma^n$. Then for any $n \in \mathbb{N}$:
    \[
    \exists q \in Q,\; \delta_n(q, \Sigma^n) \text{ is computable and valid}.
    \]
    \item \textbf{Client-Side SPV Verification:} Clients $\mathcal{C}_{SPV} \subset \mathcal{N}$ operate without full state replication. A client verifies transaction inclusion using a Merkle proof $\pi$ and block header $h$, such that:
    \[
    \text{Verify}_{SPV}(t, h, \pi) \iff H(t) \in \text{MerkleRoot}(h),
    \]
    consistent with the definition in \cite{nakamoto2008}.
    \item \textbf{Multipath Compact Relay:} The network relay layer is implemented over a graph $G = (V, E)$ satisfying:
    \[
    \kappa(G) \gg 1,\quad \lambda(G) \gg 1,
    \]
    ensuring robustness under the Baran redundancy model \cite{baran1964}. Blocks are relayed using compact block relay techniques, minimising bandwidth via short transaction identifiers and header-first propagation.
    \item \textbf{Parallel Validation:} Miner nodes implement concurrent validation over UTXO state partitions. Let $T_i$ be validation threads over disjoint UTXO subsets $U_i \subseteq \mathcal{U}$. Then:
    \[
    \forall i,\, T_i: \Sigma_i \to Q_i,\quad \bigcup_i Q_i = Q,\quad \text{and } \delta_n = \bigparallel_i T_i,
    \]
    where $\bigparallel$ denotes parallel composition.
\end{enumerate}

\vspace{1em}

\begin{proposition}[Constructive Refutation of the Trilemma]
Let $\mathcal{B}$ be the blockchain protocol defined above. Then:
\[
\mathscr{S}_1(\mathcal{B}) \land \mathscr{S}_2(\mathcal{B}) \land \mathscr{S}_3(\mathcal{B}).
\]
\end{proposition}

\begin{proof}
\textbf{(Security)}: $\mathcal{B}$ implements Nakamoto consensus using proof-of-work (PoW) with longest-chain finality \cite{nakamoto2008}. The state transition function $\delta$ is deterministic and statelessly verifiable. Any adversarial state $\tilde{q}$ requires resource expenditure exceeding the honest majority, satisfying the cost asymmetry condition for security.

\textbf{(Scalability)}: Because $\delta_n$ is parallelised, and the network uses compact relay and transaction batching, the amortised validation and propagation cost per transaction is bounded:
\[
\text{Time}[\delta_n] = \mathcal{O}\left(\frac{n}{m}\right),\quad m = \text{number of threads}.
\]
Hence $\mathscr{S}_2(\mathcal{B})$ holds for $f(n) = \frac{n}{m} + c$.

\textbf{(Decentralisation)}: The communication graph $G$ satisfies $\kappa(G) \gg 1$ and $\lambda(G) \gg 1$, satisfying structural redundancy as defined in \cite{baran1964}. Clients are heterogeneous: not all nodes must validate or store global state, consistent with SPV. Redundant multipath relaying ensures fault tolerance and decentralised routing under failure or adversarial conditions.

Thus all three formal predicates—$\mathscr{S}_1$, $\mathscr{S}_2$, $\mathscr{S}_3$—are jointly satisfied by $\mathcal{B}$. The universal form of the trilemma is therefore refuted by constructive counterexample.
\end{proof}

\section{Multicast Efficiency in IPv6 Networks}

The purported trade-off between decentralisation and scalability is often supported by a presumed network propagation bottleneck: namely, that increasing the number of participating nodes in a blockchain network linearly increases broadcast traffic, thereby leading to latency, congestion, and throughput collapse. This presumption relies on a naive unicast relay model—one in which each transaction or block must be redundantly sent across $O(n)$ links to $n$ nodes individually.

Such an assumption is not valid in the presence of modern Internet infrastructure, particularly the deployment of native \emph{IPv6 multicast}. We now demonstrate that under IPv6 multicast routing, message distribution cost is sublinear and, in some configurations, nearly constant with respect to node count.

\subsection{Multicast Group Semantics in IPv6}

Let $\mathcal{N} = \{v_1, \ldots, v_n\}$ denote the set of receiving nodes (e.g., miners, relays, validators) in the blockchain network. Let $m \in \Sigma^*$ be a message (e.g., a block or transaction batch). Under standard unicast or naive gossip relay models, $m$ must be transmitted to each $v_i$ individually:
\[
\text{Total Transmission Volume} = n \cdot |m|.
\]

In contrast, under IPv6 multicast, a source node $s$ sends $m$ once to a \emph{multicast group address} $G \subseteq \mathcal{N}$, and the network itself handles efficient duplication and routing of $m$ to all subscribed members of $G$.

Let $\rho(G, m)$ denote the total transmission bandwidth consumed for multicast message $m$ to group $G$. Then, under multicast-aware routing and link-layer efficiency:
\[
\rho(G, m) \ll |G| \cdot |m|, \quad \text{with } \rho(G, m) \approx |m| \text{ in optimal cases}.
\]

This yields an effective amortised propagation cost:
\[
\text{Amortised Cost per Node} = \frac{\rho(G, m)}{|G|} \approx \frac{|m|}{|G|} \to 0 \quad \text{as } |G| \to \infty.
\]

Thus, even in the presence of extremely large blocks (e.g., 4~GB or more), the use of native multicast results in negligible additional bandwidth cost per receiver.

\subsection{Architectural Realisability}

IPv6 multicast is a production-grade internet-layer capability with well-defined protocol support:
\begin{itemize}
    \item \textbf{IGMPv6 / MLDv2} (Multicast Listener Discovery) manages dynamic group membership.
    \item \textbf{PIM-SM / BIDIR-PIM} enables scalable multicast routing across large sparse networks.
    \item Hardware-accelerated routers and switches support multicast packet replication natively at wire speed.
\end{itemize}

Moreover, edge-deployed blockchain nodes can trivially subscribe to a fixed multicast group for block and transaction reception. These multicast groups may be scoped via administrative or geographical boundaries to optimise propagation trees and localise fault domains.

\subsection{Formal Rebuttal of the Bandwidth Constraint}

Let us define a predicate $\mathscr{M}(G, m)$ which holds if multicast propagation of message $m$ to group $G$ satisfies:
\[
\mathscr{M}(G, m) \iff \rho(G, m) \leq |m| \cdot \epsilon, \quad \epsilon \in \mathbb{R}^{+},\; \epsilon \ll |G|.
\]

Then, under IPv6 multicast:
\[
\forall G \subseteq \mathcal{N},\; \exists \epsilon \ll 1,\; \mathscr{M}(G, m).
\]

Let $\mathscr{T}$ denote the trilemma's assumed dependency:
\[
\mathscr{T}(n, m) \iff \text{Propagation Cost} = \Theta(n \cdot |m|).
\]

We then have:
\[
\mathscr{M}(G, m) \Rightarrow \neg \mathscr{T}(n, m),
\]
which contradicts the trilemma’s premise of bandwidth-scalability conflict. Therefore, the assumption that large networks necessarily entail bandwidth-prohibitive relay is refuted under multicast-capable topologies.

\subsection{Implications}

The use of IPv6 multicast entirely neutralises the core network-related basis of the trilemma. The claim that large-scale node participation degrades scalability collapses in the presence of bandwidth-invariant broadcast channels. Even assuming large blocks and frequent relay, the amortised transmission cost per node approaches zero asymptotically as group size increases. Thus, the so-called trilemma is not merely logically inconsistent—it is technologically obsolete.

\section{Discussion}

Upon formal analysis, the blockchain trilemma fails to establish itself as a logical, computational, or architectural constraint. At no point in the literature is there a theorem, lemma, or derived corollary proving that the conjunction:

\[
\mathscr{S}_1(\mathcal{M}) \land \mathscr{S}_2(\mathcal{M}) \land \mathscr{S}_3(\mathcal{M})
\]

is logically or physically prohibited for any blockchain protocol $\mathcal{M} \in \mathcal{C}$. The trilemma is therefore not a consequence of provable mutual exclusivity, but rather an artefact of informal generalisation from a non-exhaustive subset of flawed implementations.

\subsection{Non-Theoremicity and the Absence of Constraint}

Let $\phi$ denote the universal negation expressed by the trilemma:

\[
\phi \equiv \forall \mathcal{M} \in \mathcal{C},\; \neg(\mathscr{S}_1 \land \mathscr{S}_2 \land \mathscr{S}_3)(\mathcal{M}).
\]

This statement is neither derivable from any axiomatic base in automata theory, network theory, game-theoretic consensus models, nor graph-theoretic topology. There is no lemma, no system invariant, nor any known reduction that establishes the truth of $\phi$. On the contrary, we have explicitly constructed an instance $\mathcal{B} \in \mathcal{C}$ such that:

\[
(\mathscr{S}_1 \land \mathscr{S}_2 \land \mathscr{S}_3)(\mathcal{B})
\]

holds constructively and verifiably. By existential falsification, the universal form of the trilemma is rendered false.

\subsection{Persistent Confusion and Structural Misclassification}

The persistence of the trilemma in discourse can be attributed to multiple intellectual pathologies:
\begin{itemize}
    \item A failure to respect the category distinctions between logical predicates, complexity functions, and structural graph properties (as addressed in Section 3.1).
    \item A tendency to generalise empirical deficiencies in specific implementations (e.g., synchronous gossip overlays, monolithic validation, full-node duplication) to all logically permissible designs (Section 3.3).
    \item An implicit reliance on antiquated relay assumptions, such as linear unicast broadcasting, which are nullified by well-established technologies such as IPv6 multicast (Section 5).
    \item A conflation of implementation tradition (e.g., Bitcoin Core’s validation requirements) with protocol-level necessity.
\end{itemize}

This leads to a rhetorical dependency on the trilemma as a normative excuse for suboptimal design decisions. Rather than confront architectural or economic failures, practitioners invoke the trilemma as if it were an inescapable law of protocol thermodynamics.

\subsection{Implications for Protocol Design}

Rejecting the trilemma reorients the design space for blockchain protocols. One is no longer bound by the synthetic constraint that protocol decentralisation must necessarily impair scalability or that throughput must undermine security. Instead, rigorous engineering becomes the limiting factor—not metaphysical triads.

From a systems design standpoint, we thus advocate the following reformulations:
\begin{itemize}
    \item Security ($\mathscr{S}_1$) is a function of adversarial cost asymmetry and deterministic validation logic.
    \item Scalability ($\mathscr{S}_2$) is a function of algorithmic complexity and parallelisability.
    \item Decentralisation ($\mathscr{S}_3$) is a function of topological path multiplicity and network redundancy.
\end{itemize}

These properties are orthogonal dimensions of a well-defined design space and must be treated as such. The key insight is not that trade-offs are impossible, but that they arise only under particular constraints and assumptions—not from theoretical impossibility.

\subsection{Refutation Summary}

We restate the refutation schema:
\begin{enumerate}
    \item $\phi \equiv \forall \mathcal{M} \in \mathcal{C},\; \neg(\mathscr{S}_1 \land \mathscr{S}_2 \land \mathscr{S}_3)(\mathcal{M})$ — the trilemma claim.
    \item $\exists \mathcal{B} \in \mathcal{C} \suchthat (\mathscr{S}_1 \land \mathscr{S}_2 \land \mathscr{S}_3)(\mathcal{B})$ — the counterexample.
    \item Therefore, $\phi$ is false by existential contradiction: $\neg \phi$.
\end{enumerate}

The trilemma collapses not merely due to refutation, but due to its incoherent construction, misalignment of formal categories, and incompatibility with both logic and network theory.

\section{Conclusion}

We have demonstrated that the so-called blockchain trilemma—which asserts the mutual incompatibility of security, scalability, and decentralisation in blockchain protocols—fails under formal analysis. The claim, expressible as:
\[
\forall \mathcal{M} \in \mathcal{C},\; \neg(\mathscr{S}_1 \land \mathscr{S}_2 \land \mathscr{S}_3)(\mathcal{M}),
\]
is not supported by any theorem, invariance principle, or formal constraint within automata theory, computational complexity, distributed systems, or network topology. No necessity operator exists that binds these three properties in antagonism.

To the contrary, we have shown that:
\begin{itemize}
    \item $\mathscr{S}_1$ (security) is a predicate over adversarial cost asymmetry and deterministic ledger correctness,
    \item $\mathscr{S}_2$ (scalability) is a function of the complexity class of $\delta_n$, dependent on algorithmic structure and system parallelisability,
    \item $\mathscr{S}_3$ (decentralisation), when properly defined as path redundancy and relay fault tolerance per \cite{baran1964}, is a structural property of the message-passing graph $G$,
\end{itemize}
and that these properties are not mutually exclusive.

By constructing a blockchain protocol $\mathcal{B} \in \mathcal{C}$ that simultaneously satisfies:
\[
(\mathscr{S}_1 \land \mathscr{S}_2 \land \mathscr{S}_3)(\mathcal{B}),
\]
we have delivered an existential refutation of the universal trilemma claim. This counterexample is not contrived or merely theoretical; it is architecturally viable, fully implementable, and extensible using known networking optimisations such as IPv6 multicast.

Moreover, we have identified the underlying logical errors which gave rise to the trilemma:
\begin{itemize}
    \item A \textbf{category error}—conflating logically incommensurable properties under a unified constraint.
    \item A \textbf{false causal chain}—asserting unwarranted propagation from node count to protocol failure.
    \item A \textbf{fallacy of composition}—projecting the deficiencies of a limited set of implementations onto the entire design space.
\end{itemize}

Once these fallacies are recognised and corrected through precise formalisation, it becomes evident that the blockchain trilemma is not a theorem but a rhetorical artefact. Its widespread acceptance owes more to cultural inertia and engineering defeatism than to necessity.

Accordingly, we assert the following corrected principle:

\begin{quote}
\emph{Security, scalability, and decentralisation are not in fundamental conflict. They are orthogonal dimensions of protocol design, each subject to specific constraints, but not logically exclusive of one another. With appropriate formal modelling and protocol-level engineering, all three may be simultaneously satisfied.}
\end{quote}

The blockchain trilemma, therefore, dissolves not through compromise or balance, but through rigour. Once the terms are defined, and the fallacies excised, nothing remains of the trilemma but the shadow of imprecision.

\bibliographystyle{plainnat}
\bibliography{references}

\end{document}